\documentclass[pre,aps,preprint,superscriptaddress,showpacs]{revtex4}
\usepackage{amsmath,amsbsy,amssymb,graphicx}
\usepackage{color}
\usepackage{psfrag}

\begin{document}

\title{Hierarchical models of rigidity percolation}
\author{J.~Barr{\'e}}

\affiliation{ Laboratoire J. A. Dieudonn\'e, UMR CNRS 6621,
Universit\'e de Nice Sophia-Antipolis, Parc Valrose, F-06108 Nice
Cedex 02, France.}

\date{\today}
            
\begin{abstract}

  We introduce models of generic rigidity percolation in two dimensions on
  hierarchical networks, and solve them exactly by means of a
  renormalization transformation. We then study how the possibility
  for the network to self organize in order to avoid stressed bonds
  may change the phase diagram. In contrast to what happens on random
  graphs and in some recent numerical studies at zero temperature, we
  do not find a true intermediate phase separating the usual rigid and
  floppy ones.

\end{abstract}                       


\maketitle

\section{Introduction}     
Consider a structure made of sites, connected by links. Each link
imposes a constraint, by prescribing the distance between the two
sites it connects: if the actual distance between the two sites is
different from the prescription, there is an associated energy cost.
Rigidity theory deals with properties associated with the structure's
topology, which do not depend on the physical nature of the
constraints, nor on the precise form of the energy cost. Typical
questions asked by rigidity theory are: how many degrees of freedom
are left in the system? Is there a macroscopic cluster of sites
rigidly connected one with the others?  By contrast, questions related
to the elastic properties of the structure do depend on the
specification of the constraints. When the number of links in the
structure is increased, the phenomenology is as follows. For a small
enough number of links, that is a small mean connectivity of the
structure, there are many more degrees of freedom than constraints;
there is no macroscopic rigid cluster, and many degrees of freedom are
left in the system: the system is said to be floppy. At large mean
connectivity, there are many more constraints than degrees of freedom;
there is a macroscopic rigid cluster, and many constraints cannot be
satisfied: the system is said to be stressed rigid. In between these
two phases takes place the rigidity percolation transition.

Despite its clearly mechanical origin, the problem of rigidity
percolation has also attracted attention in the last 30 years because
of its applications in understanding the properties of network forming
glasses, like GeSe or GeAsSe alloys~\cite{Phillips79,Thorpe83}. In
this case, the atomic bonds may be considered as constraints.
However, contrary to standard rigidity percolation as presented in the
first paragraph, angles between adjacent bonds also need to be
considered as constraints, in addition to bond lengths. In the
following, we will concentrate on central force rigidity (that is when
only bond lengths are considered as constraints), in two dimensions.
Beyond the applications to glasses, models of cross linking stiff
fibers forming random networks were shown to fall in the same
universality class as central force 2D rigidity~\cite{LatvaPRE01};
this type of system has been used to model network forming
proteins~\cite{HeadPRL03}.

A decisive theoretical progress was made in the 90's for
\emph{generic}\footnote{an embedding of a graph in $\mathbb{R}^2$ is
  said to be generic if sites and bonds do not present any "special"
  property, like two parallel bonds, or three aligned sites. For a
  clear and rigorous presentation of generic 2D rigidity, see for
  instance~\cite{Hendrickson}.} 2D rigidity, with the introduction of
combinatorial algorithms~\cite{PebbleJacobs,PebbleMoukarzel}, based on
Laman's theorem~\cite{Laman}. These new algorithms allow for the study
of much larger samples than before, as well as more precise estimates
around the critical point~\cite{PebblePRE}. In particular, for
rigidity percolation on regular lattices, the scenario of a second
order phase transition in a different universality class than ordinary
percolation seems favored by the numerics, although there has been
some debate on the subject~\cite{MoukarzelRBM99}.

From the analytical point of view however, progresses have been slow.
A field theoretical attempt by Obukhov~\cite{Obukhov} predicts a first
order phase transition in 2D, which, when compared with the numerics,
seems to be a non generic feature. Some insight came from rigidity
models exactly solved on trees and various types of random graphs,
with locally tree-like
topology~\cite{DuxburyPRE97,ThorpeBook99,MoukarzelRBM99,ChubinskyThesis,BBLSjsp}.
However, in these types of models, the rigidity percolation transition
is also usually first order. This casts doubt on their usefulness to
understand generic 2D rigidity. There has been also earlier attempts
to study rigidity percolation through Position Space Renormalization
Group~\cite{Sahimi85,Sahimi92}; based on small renormalization cells,
the associated predictions for the critical exponents are unprecise.
To this date, we are not aware of any precise analytical prediction
for the critical exponents of 2D generic rigidity percolation.

A few years ago, Thorpe et al. have opened a new research direction by
introducing the notion of network self-organization in rigidity
percolation models~\cite{ThorpeJNCS00}. This is a natural idea in the
context of network glasses modeling: constraints that are not
satisfied create stress and bear an energy cost; the network should
then self-organize, i.e. tend to modify its structure, in order to
minimize this energy cost.  In~\cite{ThorpeJNCS00}, numerical
simulations allowing self organization point to the existence of a new
phase in between the usual floppy and rigid ones. As a function of the
mean connectivity, the phase diagram would then show two phase
transitions instead of one.  Around the same time, several experiments
on network glasses drew a picture compatible with this predicted
phenomenology~\cite{BoolchandPRB00,BoolchandEPL00,BoolchandJNCS01}. A
problem of the original simulations of~\cite{ThorpeJNCS00} was their
strongly out of equilibrium character; some theoretical studies
improved on this, and confirmed the ``three phases''
picture~\cite{MicoulautEPL02,MicoulautPhillips}.
In~\cite{BBLS_prl,RivoireBarre}, a class of self-organizing rigidity
percolation models is exactly solved, and shows three phases,
separated by two true thermodynamic phase transitions. However, these
exactly solvable models are based on random networks, which have a
locally tree like topology, without small loops. This strongly
influences the critical properties of the rigidity transition; in
particular, as already mentioned above, rigidity percolation is
continuous for a regular 2D triangular network, and it is first order
in random networks. The simulations of
refs.~\cite{ChubynskyBriere06,ChubynskyBriere06b}, performed at zero
temperature on a triangular lattice, seem to confirm the phenomenology
suggested by the locally tree like random networks studies, and show
that the entire intermediate phase has a critical nature. Sartbaeva et
al. in~\cite{Sartbaeva07} discuss the influence of local structures on
the existence and width of the intermediate phase in more realistic
network glasses models. Despite these works, the question of whether
self-organized rigidity percolation on two dimensional networks at
finite temperature displays a true "intermediate phase" is still open.
We conclude this paragraph by mentioning recent reviews on the
subject~\cite{MicoulautJNCS07,Chubynsky08}.
 
The goal of this paper is twofold. First, we introduce a new class of
solvable models of 2D rigidity, on hierarchical lattices. These
lattices, in contrast with random graphs are not locally tree like,
and possess many small loops. We may then expect for these lattices a
phenomenology closer to generic 2D rigidity than random graphs. It is
known that models defined on hierarchical lattices may be exactly
solvable, using renormalization
transformations~\cite{Ostlund,Kaufman81,Kaufman82}; we will use the
same ideas in the context of rigidity percolation. Second, in the
spirit of~\cite{BBLS_prl} for random networks, we add to the
hierarchical networks the possibility of self-organization. This
provides a new class of exactly solvable models, genuinely different
from the random graphs of~\cite{BBLS_prl,RivoireBarre}, where it is
possible to investigate the existence of an ``intermediate phase''.

The paper is organized as follows. In
section~\ref{sec:generic_rigidity}, we summarize the combinatorial
approach to 2D rigidity, which will be necessary in the following.
After introducing, in section~\ref{sec:hierarchical}, the class of
hierarchical lattices we will be interested in, we solve our model and
study its critical properties in section~\ref{sec:solution}. We then
introduce the possibility of self-organization for the hierarchical
networks, and solve the associated model, in
section~\ref{sec:adaptive}; we discuss the implications on the
existence of an "intermediate phase", between the usual floppy and
stressed rigid ones. Section~\ref{sec:conclusion} is devoted to
discussion and perspectives.

\section{Generic 2D rigidity}
\label{sec:generic_rigidity}
\begin{figure}
\includegraphics[scale=0.4]{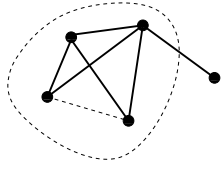}
\caption{An example of a graph with $N=5$ sites. Before the addition 
of the dashed bond, there are $M=6$ bonds, none of which is redundant. 
The number of floppy modes is~$4$: two global translations, one global 
rotation, and the additional rotation of the one-coordinated site. 
The dashed bond is redundant: the subgraph isolated by the dashed curve 
has $N_{sub}=4$ sites and $M_{sub}=6>2N_{sub}-3$ bonds. The whole graph with 
the dashed bond also has $4$ floppy modes, in agreement with 
Eq.~(\ref{eq:flop}).
\label{Fig:Laman_example}}
\end{figure}

An intuitive approach to rigidity percolation, apparently due to
Maxwell, consists in counting degrees of freedom and constraints. In
two dimensions, each site has two degrees of freedom; each link brings
one constraint. In the end, a certain number of degrees of freedom are
left.  They correspond to unconstrained motions of the sites, and are
usually called "floppy modes". If one thinks of each link as a spring,
with a given rest length, floppy modes are motions of the sites that
do not change any spring length, and thus do not cost any energy. 
Any network embedded in two dimensions necessarily has at least three
floppy modes, corresponding to the two global translations and the
global rotation in the plane. The network, or a subgraph of the
network, is said to be rigid if it has no internal degree of freedom,
besides these three. In a first approximation, a network with $N$
sites and $M$ links should have
\begin{equation}
N_{flop}=\max(3,2N-M)
\label{eq:max} 
\end{equation}
degrees of freedom left, and be rigid if $2N-M\leq 3$. This simple
computation is obviously wrong in general, as shown on
Fig.~\ref{Fig:Laman_example}. The reason is that when a new bond is
added, it may be ``redundant'': then the constraint associated with
this new bond cannot be satisfied and its addition does not remove any
degree of freedom. The exact formula giving the number of floppy modes
of a network is
\begin{equation}
N_{flop}=2N-M+N_{red}~,
\label{eq:flop}
\end{equation}
where $N_{red}$ is the number of redundant bonds. The problem is then
reduced to the estimation of $N_{red}$. The simplest form of Maxwell
counting given by Eq.~(\ref{eq:max}) assumes that new bonds added in
the network first exhaust the degrees of freedom available before
starting to become redundant. It is not correct, but it turns out that
a generalization of this idea is actually exact~\cite{Laman}. It may
be formulated as follows. Consider a graph, and add a new bond to it.
It is redundant if and only if there exists a subgraph with $N_{sub}$
sites and $M_{sub}$ bonds, containing this new bond, such that
$2N_{sub}-M_{sub}<3$. This combinatorial formulation is at the root of
the ``pebble game'', a powerful algorithm to analyze rigidity
percolation on two dimensional
networks~\cite{PebbleJacobs,PebbleMoukarzel}.

\section{Hierarchical lattices}
\label{sec:hierarchical}
\begin{figure}
\psfrag{2}{\large \hskip-.2truecm $1)$}
\psfrag{1}{\large \hskip-.2truecm $2)$}
\psfrag{3}{\large \hskip-.2truecm $3)$}
\psfrag{4}{\large \hskip-.2truecm $4)$}
\includegraphics[scale=0.4]{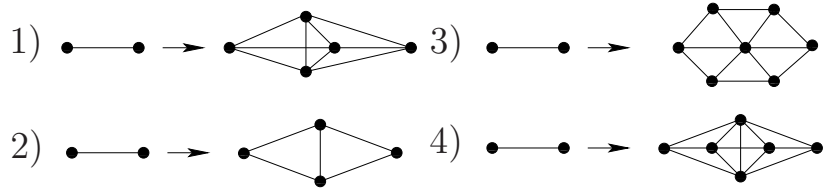}
\caption{Four examples of hierarchical graphs. On the left, the graphs
  at step $t=0$, with two sites and one bond; on the right, the graphs
  at $t=1$. We call the $t=1$ graph the ``elementary cell''. $1)$ is
  an elementary cell with $5$ sites and $9$ bonds. This is the
  hierarchical graph we will mostly consider in the following. $2)$,
  $3)$ and $4)$ are elementary cells with respectively $4$ sites and
  $5$ bonds, $7$ sites and $12$ bonds, and $6$ sites and $12$ bonds.
  At step $t$, each bond of step $t-1$ is replaced by a subgraph
  identical to the $t=1$ graph.
\label{Fig:hierarchical_examples}}
\end{figure}

We will consider the following type of hierarchical graphs. We start
from two sites, connected by one bond. The graph is then constructed
iteratively; at each step, all bonds are replaced by a given
elementary cell. Four examples of elementary cells,
  corresponding to four examples of hierarchical graphs, are given on
Fig.~\ref{Fig:hierarchical_examples}.
In each case, it is easy to obtain a recursion relation for the number
of sites $N_{t}$ and bonds $B_{t}$ of the graphs after $t$ iterations.
We obtain respectively, for the four graphs of
Fig.~\ref{Fig:hierarchical_examples}
\begin{eqnarray}
B_{t+1}^{(1)}=9B_{t}^{(1)} &;& N_{t+1}^{(1)}= N_t^{(1)}+3B_t^{(1)}\\
B_{t+1}^{(2)}=5B_{t}^{(2)} &;& N_{t+1}^{(2)}= N_t^{(2)}+2B_t^{(2)}\\
B_{t+1}^{(3)}=12B_{t}^{(3)} &;& N_{t+1}^{(3)}=N_t^{(3)}+5B_t^{(3)}\\
B_{t+1}^{(4)}=12B_{t}^{(4)} &;& N_{t+1}^{(4)}=N_t^{(4)}+4B_t^{(4)}~.
\end{eqnarray}
Solving these recursion relations yields:
\begin{eqnarray}
B_t^{(1)}=9^t &;& N_t^{(1)}=\frac{3\times 9^t+13}{8} \\
B_t^{(2)}=5^t &;& N_t^{(2)}=\frac{5^t+3}{2} \\
B_t^{(3)}=12^t &;& N_t^{(3)}=\frac{5\times 12^t+17}{11}\\
B_t^{(4)}=12^t &;& N_t^{(4)}=\frac{4\times 12^t+18}{11}~.
\end{eqnarray}
To model the rigidity percolation phenomenon, we now assume that each
bond is effectively present in the graph with probability $p$; thus,
the number of occupied bonds is, for large graphs, $M_{t} \simeq
pB_{t}$. The Maxwell counting procedure yields an estimate of the
threshold for rigidity percolation: the approximate critical
probability $p_{Max}$ corresponds to a ratio between bonds and sites
$M_t/N_t=2$. For graph on Fig.~\ref{Fig:hierarchical_examples}.2, we
find $p_{Max}^{(2)}=1$; any $p<1$ should then render this graph
floppy, and there is no true rigidity percolation.  We thus dismiss
this example in the following. For the graphs on
Figs.~\ref{Fig:hierarchical_examples}.1,~\ref{Fig:hierarchical_examples}.3
and~\ref{Fig:hierarchical_examples}.4, we find respectively
$p_{Max}^{(1)}=3/4$, $p_{Max}^{(3)}=10/11$ and $p_{Max}^{(4)}=8/11$.
All three graphs should then present a rigidity transition at $p<1$.
We will mainly concentrate in the following on graph $1)$ for the sake
of simplicity.

\section{Solution of the hierarchical models}
\label{sec:solution}
The goal is to compute a recursion relation relating $N_{t}^{red}(p)$,
the number of redundant bonds of a graph iterated $t$ times, with a
bond occupation probability $p$, and $N_{t-1}^{red}(p')$, the number
of redundant bonds of a graph iterated $t-1$ times, with a bond
occupation probability $p'$.  This will be possible because of the
peculiar structure of the network~\cite{Kaufman81}.

\begin{figure}
\includegraphics[scale=0.7]{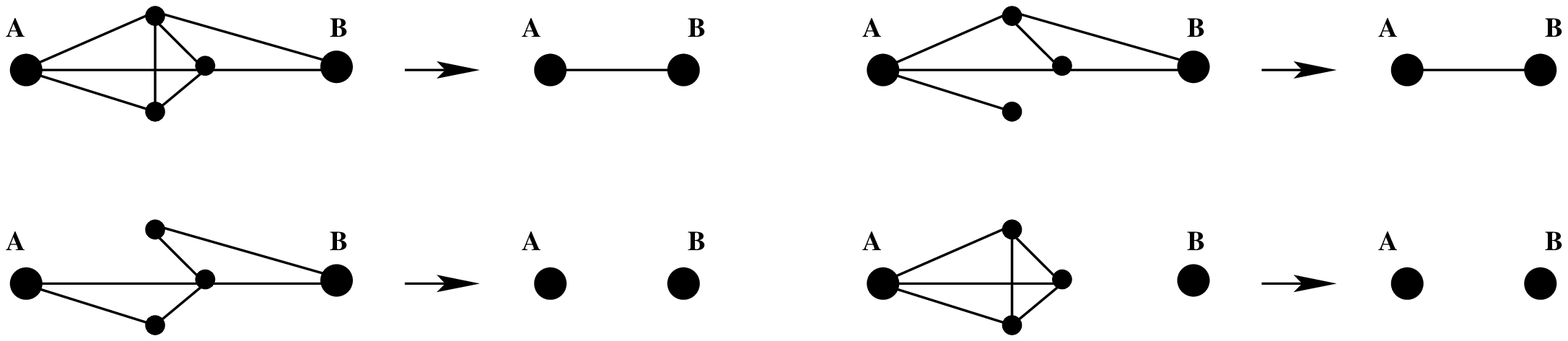}
\caption{
Renormalization of an elementary cell, and counting of redundant 
bonds. Top left: sites $A$ and $B$ are rigidly connected, so 
that the elementary cell is renormalized into a bond; there is one 
redundant bond. 
Top right: sites $A$ and $B$ are rigidly connected, no redundant bond.
Bottom left: sites $A$ and $B$ are not rigidly connected, no redundant bond. 
Bottom right: sites $A$ and $B$ are not rigidly connected, one redundant bond. 
\label{Fig:renorm}}
\end{figure}

The procedure is somewhat reciprocal to the construction of the
hierarchical network. Consider a graph iterated $t$ times, with a bond
probability $p$. Consider two sites $A$ and $B$ separated by one
elementary cell, see Fig.~\ref{Fig:renorm}. If they are rigidly
connected through this cell, we replace the whole cell by a single
bond. If they are not, we delete the whole cell. Repeating this
procedure for all elementary cells, we construct a new graph, iterated
$t-1$ times, with a bond occupation probability $p'=\varphi(p)$. The
function $\varphi(p)$ is constructed by counting all configurations
leading to a rigid connection between $A$ and $B$. This process is
detailed on Table~\ref{tab1}, for the graph on 
Fig.~\ref{Fig:hierarchical_examples}.1.
\begin{table}[ht]
\begin{tabular}{|c|c|c|c|}
\hline
nbr bonds & rigid connection  & nbr redundant bonds & nbr configurations\\
\hline
9 & yes & 2 & 1\\
8 & yes & 1 & 9\\
7 & yes & 0 & 30\\
7 & no & 1 & 6\\
6 & yes & 0 & 12\\
6 & no & 0 & 70\\
6 & no & 1 & 2\\
5 & yes & 0 & 3\\
5 & no & 0 & 123\\
4 & no & 0 & 126\\
3 & no & 0 & 84\\
2 & no & 0 & 36\\
1 & no & 0 & 9\\
0 & no & 0 & 1\\
\hline
\end{tabular}

\caption{Enumeration of all types of possible configurations for an 
  elementary cell of
  the graph on Fig.~\ref{Fig:hierarchical_examples}.1. 
  There are for instance $12$ configurations with $6$ bonds, providing a rigid 
  connection across the cell, and without redundant bond. Some examples are 
  given on Fig.~\ref{Fig:renorm}.}
\label{tab1}
\end{table}

Collecting the different contributions, we obtain
\begin{equation}
\varphi(p)=p^9+9p^8q+30p^7q^2+12p^6q^3+3p^5q^4~,
\end{equation}
where $q=1-p$.
We assume that the number of redundant bonds is an extensive quantity;
neglecting subdominant contributions and probabilistic fluctuations,
we then write
\begin{equation}
N_{t}^{red}(p)=N_{t}r(p)~,
\end{equation}
and our goal is to compute $r(p)$. Analyzing one renormalization step, we write
\begin{equation}
N_{t}^{red}(p)=N_{t-1}^{red}(\varphi(p))+N'_{red}(t,p)
\label{eq:nred}
\end{equation}
where $N'_{red}(t,p)$ represents all the redundant bonds that where
suppressed in the renormalization process. This formula needs some
justification. Consider the empty (without bonds) graph at step $t$,
and add the bonds one by one, checking each time if the newly added
bond is redundant or not; this is actually the numerical strategy
implemented in the ``pebble game'' algorithm. If the added bond is
redundant, it is discarded. Thus, during the process, the constructed
graph has at most one redundant bond. If an added bond is redundant,
it is possible to find a subgraph $S$, containing this new bond, with
$N_{sub}$ sites and $M_{sub}=2N_{sub}-2$ bonds. We choose $S$ 
as small as possible.
There are now two possibilities. Either $S$ is included in one
elementary cell; in this case the redundant bond is included in the
$N'_{red}(t,p)$ contribution. Or $S$ is not included in an elementary
cell. It is easy to see that no floppy elementary cell can be included
in $S$, so that the newly added bond necessarily ``rigidifies'' the
elementary cell it belongs to. In addition, if $S$ contains a bond included 
in one elementary cells, $S$ contains both sites which are end points of 
the cell, and the cell is rigid. $S$, which is included in the
graph at step $t$ will then be renormalized in a natural way in a
subgraph of the graph at step $t-1$.  The redundant constraint is thus
included in the $N_{t-1}^{red}(\varphi(p))$ contribution to
Eq.~(\ref{eq:nred}). It is worth mentioning that both the hierarchical
nature of the graph and the convenient combinatorial description of 2D
rigidity outlined in Section~\ref{sec:generic_rigidity} are necessary
for this step. In particular, it is not clear how to construct exactly
solvable hierarchical models of 3D rigidity.

For a given bond probability $p$, we want to compute $r_{0}(p)$, the
mean number of redundant bonds inside one elementary cell.  We use
Table~\ref{tab1} to enumerate all possibilities, and we obtain the
following expression for $r_{0}(p)$:
\begin{equation}
\label{eq:r0}
r_0(p)=2p^9+9p^8q+6p^7q^2+2p^6q^3
\end{equation}

In one renormalization step, a large number of elementary cells
$B_{t-1}$ is renormalized. Thus $N'_{red}(t,p)\simeq B_{t-1}r_{0}(p)$.
Finally, the recursion relation reads
\begin{equation}
N_{t} r(p)=N_{t-1}r(\varphi(p))+B_{t-1}r_{0}(p)
\end{equation}
Using the relations $B_t=B^t$ and $B_t/N_t=x_{0}$, and simplifying by
$N_{t}$, this yields
\begin{equation}
r(p)=\frac{1}{B}r(\varphi(p))+\tilde{r}_{0}(p)~,
\label{eq:rec}
\end{equation}
with $\tilde{r}_{0}(p)=x_{0}r_{0}(p)/B$. For our main example, represented on 
Fig.~\ref{Fig:renorm}, $B=9$ and $x_{0}=8/3$.

From Eq.~(\ref{eq:flop}), we see that for given numbers of sites and
bonds, the number of floppy modes is linearly related to the number of
redundant bonds.  Since it has been noticed~\cite{MoukarzelRBM99} that
the number of floppy modes plays the role of a "free energy" in
rigidity percolation models, we remark that the computation of $r(p)$
is analogous to the free energy computation of ref.~\cite{Kaufman82},
for spin models on hierarchical lattices.

\begin{figure}
\psfrag{f}{\Large $\varphi$}
\psfrag{x}{\Large $p$}
\includegraphics[width=12cm,height=8cm]{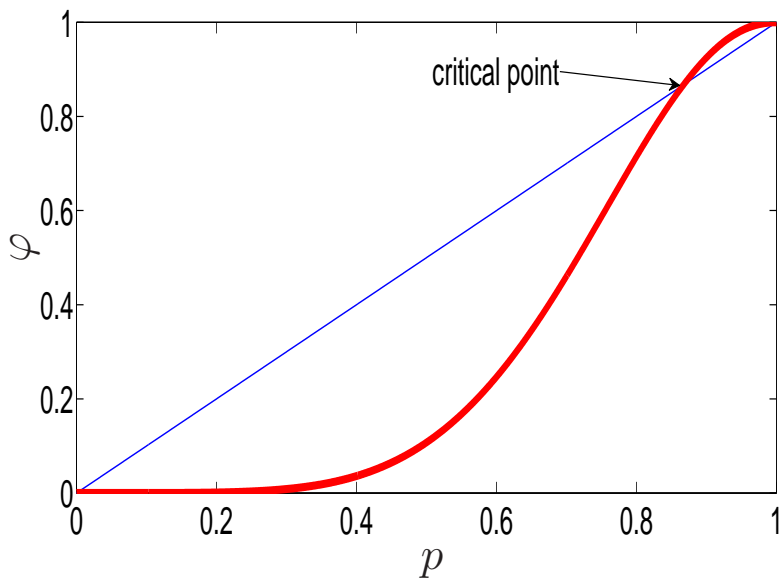}
\caption{(Color online) The red curve $y=\varphi(p)$ (thick line), together 
with the blue $y=p$ straight line (thin line). The critical point defined by 
$\varphi(p^*)=p^*$ is indicated by an arrow.
\label{fig:phi}}
\end{figure}

The function $\varphi$ is plotted on Fig.~\ref{fig:phi}. It has three
fixed points; $p=0$ and $p=1$ are stable and correspond to the floppy
and rigid phases respectively. $p=p^*\simeq 0.867$ is unstable and is
the critical point. We note $\varphi^{(k)}$ the $k$th iterate of
$\varphi$.  Then for all $p<p^*$ (resp. $p>p^*$), $\varphi^{(n)}(p)$
tends to $0$ (resp. $1$). The quantity $a=d\varphi/dp(p=p^*)\simeq
1.97$ will play an important role. We note that if we had used the
hierarchical network on Fig.~\ref{Fig:hierarchical_examples}.2, we
would have obtained a function $\varphi$ with only two fixed points:
$p=0$, stable, and $p=1$, unstable. Thus, in this case, any $p<1$
yields a floppy network at large scales. A similar analysis for
  the graphs on Figs.~\ref{Fig:hierarchical_examples}.3
  and~\ref{Fig:hierarchical_examples}.4 yields respectively
  $p^*=0.975$, $p^*=0.853$ for the critical densities, and $a=1.86$,
  $a=2.07$ for the $\varphi$ derivative at criticality. We note that
  in all three cases, we find $p^*>p_{Max}$.

Iterating $n$ times Eq.~(\ref{eq:rec}), one obtains
\begin{eqnarray}
  r(p)&=&\frac{1}{B^n} r(\varphi^{(n)}(p))+\frac{1}{B^{n-1}}\tilde{r}_0(\varphi^{(n-1)}(p))+\ldots+ \nonumber \\
&&+\frac{1}{B}\tilde{r}_0(\varphi(p))+\tilde{r}_0(p)
\end{eqnarray}
As $B>1$ and $\tilde{r}_0$ is bounded, $r$ may then be expressed as an
infinite series:
\begin{equation}
r(p)=\sum_{k=0}^{\infty}\frac{1}{B^{k}}\tilde{r}_0(\varphi^{(k)}(p))
\end{equation}
As the $k$ times iterated function $\varphi^{(k)}(p)$ tends pointwise
to $0$ when $p<p^*$, and to $1$ when $p>p^*$, the above series is
infinitely differentiable, except when $p=p^*$, the unstable fixed
point of $\varphi$.  Looking for a power law behavior of $r$ close to
$p^*$ 
\begin{equation}
r(p)= Cte |p-p^*|^\tau + \mbox{regular part}~,
\end{equation}
one finds $\tau=\ln B/\ln a$. For our main example, we find
$\tau=3.24$; this corresponds to a critical exponent $\alpha=-1.24$,
where we have used the standard notation for the critical exponent
associated with the free energy singularity.  This means that the
derivatives of $r(p)$ are all continuous up to third order, while the
fourth derivative is discontinuous. This does not compare well to the
numerical results on triangular networks~\cite{PebblePRE}, which find
$\alpha \simeq -0.48$. Networks on
Fig.~\ref{Fig:hierarchical_examples}.3 and
Fig.~\ref{Fig:hierarchical_examples}.4 yield respectively $\tau=4.01$
and $\tau=3.41$, which corresponds to $\alpha=-2.01$ and
$\alpha=-1.41$. Thus, there are important fluctuations in the value of
the critical exponent between two different hierarchical networks. We
note however that the tendency to a very weak rigidity percolation
transition on these hierarchical networks seems to be general. A
natural idea to improve the agreement between these analytical results
on hierarchical networks and the numerics on triangular lattices would
be to consider larger renormalization cells. The previous remarks cast
doubt on this strategy. Another interesting extension would be to
study other critical properties, like the size of the largest rigid
cluster, or the largest stressed cluster.

\section{Adaptive networks}
\label{sec:adaptive}

\subsection{Introduction of an energy}

We consider now the possibility for the network to modify its
structure, in order to avoid stress. This step was first taken
in~\cite{ThorpeJNCS00}, in an attempt to better model network glasses;
in this context, it is natural to assume that the connectivity
structure of the network may adapt itself to a certain extent to avoid
stressed bonds. Following~\cite{BBLS_prl}, we introduce an energy cost
for each constraint which is not satisfied; at any finite temperature,
this cost competes with the entropic cost of reorganizing the network.
In this setting, standard rigidity percolation, as studied up to now,
corresponds to the infinite temperature limit, where the energy cost is
negligible.

We have to define an energy for any configuration of the network.  The
simplest choice is to count the number of redundant bonds, and to
attribute the same amount of energy to each of
them~\cite{BBLS_prl,RivoireBarre}.

We consider a network where a proportion $x$ of the bonds is present.
The partition function for our model, on a hierarchical network
iterated $t$ times, reads:
\begin{equation}
  Z_{t}(x,\beta)=\sum_{\{l_{j}\},\sum l_{j}=xB_{t}} e^{-\beta N_{red}(\{l_{j}\})},
\end{equation}
where we note $l_{j}=1$ if bond $j$ is present, $l_{j}=0$ otherwise;
$\sum_{\{l_{j}\}}$ is the sum over all possible configurations of the network.
Here, it is restricted to the configurations with the prescribed fraction of
occupied bonds, $x$.

\begin{figure}
\psfrag{x}{\Large $x$}
\psfrag{p}{\Large $p$}
\includegraphics[width=12cm,height=8cm]{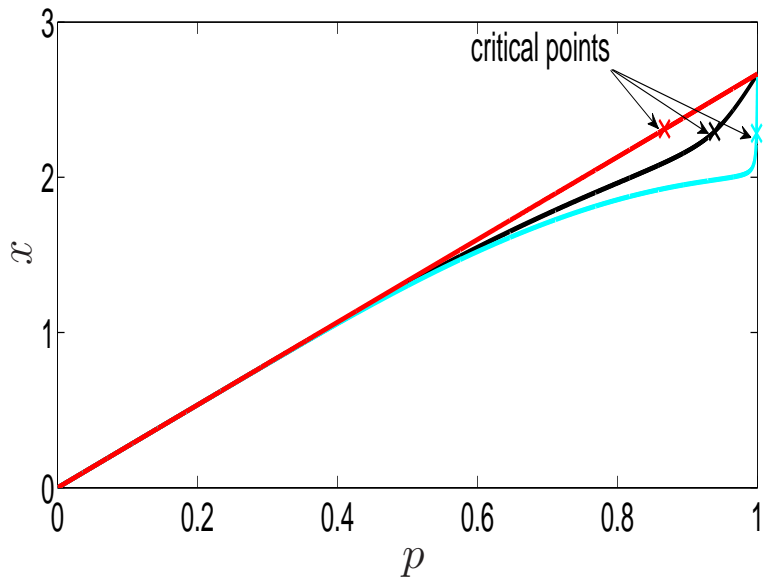}
\caption{(Color online) The mean connectivity of the network $x$ as a
  function of $p$.  From left to right, $\beta=0,~1,~5$ (respectively
  red, black and light blue curves). The critical points are also
  indicated. \label{Fig:xp}}
\end{figure}

\begin{figure}
\psfrag{x}{\Large $x$}
\includegraphics[width=12cm,height=8cm]{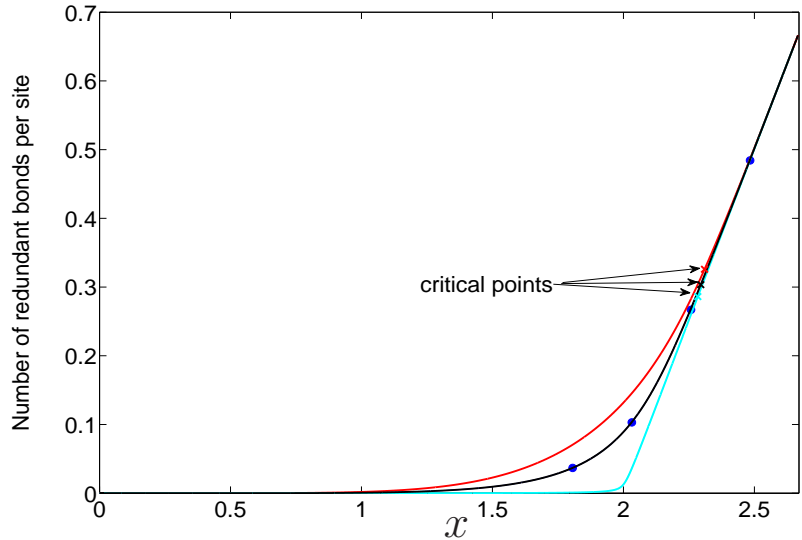}
\caption{(Color online) The number of redundant per site is plotted as
  a function of the mean connectivity $x$, for $\beta=0,~1,~5$ (from
  left to right, respectively red, black and light blue curves).  The
  crosses indicate the critical points; the filled circles are results
  of Metropolis Monte Carlo simulations, for $\beta=1$, using the
  pebble game algorithm, on hierarchical graphs with $t=5$
  ($22145$~sites). They are in perfect agreement with the analytical
  results. \label{Fig:nred}}
\end{figure}

To proceed, it is useful to relax the constraint on the number of
bonds. We introduce a "grand canonical" statistical ensemble, whose
partition sum reads:
\begin{equation}
\Omega_{t}(p,\beta)=\sum_{\{l_{j}\}} p^{\sum l_{j}}q^{B_{t}-\sum l_{j}}e^{-\beta N_{red}(\{l_{j}\})}~,
\end{equation}
where $p$ is the a priori probability that a bond is present, and
$q=1-p$. $\sum l_{j}$ is the total number of bonds in the network in 
configuration $\{l_{j}\}$. We
recall that $B_{t}$ is the maximum number of bonds.
Because of the energetic term which biases the probability
distribution towards non stressed configurations, the actual number of
bonds present in the network is different from $pB_{t}$, except in the
infinite temperature limit.  We may rewrite the partition sum as
\begin{equation}
  \Omega_{t}(p,\beta)=\left(\frac{e^{\lambda}}{1+e^{\lambda}}\right)^{B_{t}}
\sum_{\{l_{j}\}} e^{-\lambda \sum l_{j}} e^{-\beta N_{red}(\{l_{j}\})},
\label{eq:eqOmega2}
\end{equation}
where 
\begin{equation}
\label{eq:lambda}
\lambda=\ln \left(\frac{1-p}{p}\right)
\end{equation}
We define $\omega(p,\beta)=\ln \Omega_{t}(p,\beta)/ N_{t}$. Generalizing
the renormalization transformation of the previous section, it is
possible to compute exactly $\omega(p,\beta)$. We consider one
elementary cell, and want to compute $\varphi(p,\beta)$, the
probability that this elementary cell is renormalized as a bond. The
calculation is the same as in section~\ref{sec:solution}, except that
we have to attribute a weight $e^{-n\beta}$ instead of $1$ to
configurations with $n$ redundant bonds. For the elementary cell of
our chief example, Fig.~\ref{Fig:hierarchical_examples}.1, we use once again
Table~\ref{tab1}, to obtain
\begin{equation}
  \varphi(p,\beta)=\frac{\varphi_{1}(p,\beta)}{\varphi_{1}(p,\beta)+\varphi_{0}(p,\beta)},
\end{equation}
where $\varphi_{1}$ collects the contributions of the configurations
where the end sites of the cell are rigidly connected, and
$\varphi_{0}$ collects the contributions of the configurations where
the end sites of the cell are not rigidly connected. Division by
$\varphi_{1}+\varphi_{0}$ is for normalization. We have
\begin{eqnarray}
\varphi_{1}(p,\beta)&=&p^9e^{-2\beta}+9p^{8}qe^{-\beta}+30p^7q^2+12p^6q^3+
3p^5q^4 \nonumber \\
\varphi_{0}(p,\beta)&=&6p^7q^2e^{-\beta}+2p^{6}q^3e^{-\beta}+70p^6q^3+
123p^5q^4 \nonumber \\
&&+126p^4q^5+84p^3q^6+36p^2q^7+9pq^8+q^9
 \end{eqnarray}
\begin{figure}
\psfrag{f}{\Large $\varphi$}
\psfrag{p}{\Large $p$}
\includegraphics[width=12cm,height=8cm]{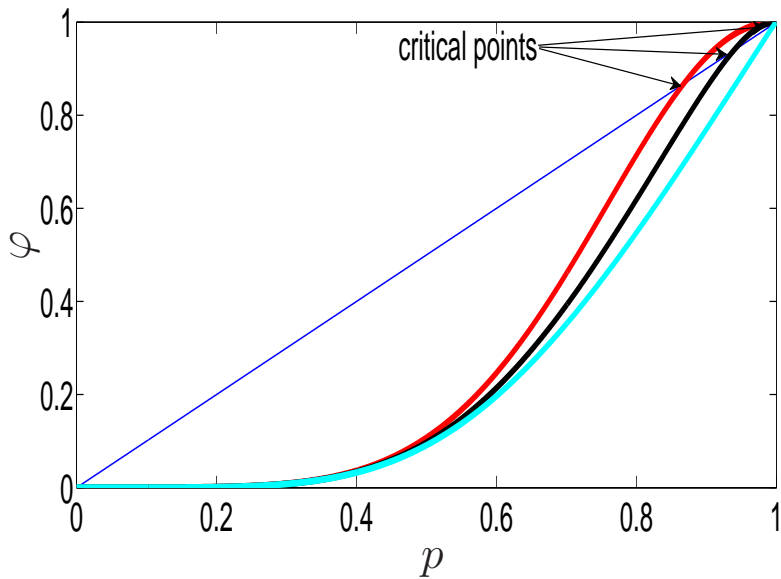}
\caption{(Color online) Examples of functions $\varphi(p,\beta)$, as a
  function of $p$; from left to right, $\beta=0,~1,~5$ (thick lines,
  respectively red, black and light blue curves). The intersection with the
  $y=p$ straight line (blue thin line) is the critical point. A low
  temperature shifts the critical point towards high $p$ values.
  \label{Fig:phiT}}
\end{figure}
 Some curves $\varphi(p,\beta)$ are drawn for various values of
 $\beta$ on Fig.~\ref{Fig:phiT}. 
We also define
\begin{equation}
c(p,\beta)=\varphi_{1}(p,\beta)+\varphi_{0}(p,\beta)
\end{equation}
Then, following the same reasoning as above, we derive the recursion relation
\begin{equation}
\Omega_{t}(p,\beta)=c(p,\beta)^{B_{t-1}}\Omega_{t-1}(\varphi(p,\beta),\beta)
\end{equation}
Taking the logarithm:
\begin{equation}
N_{t}\omega(p,\beta)=B_{t-1}\ln c(p,\beta) +N_{t-1}\omega(\varphi(p,\beta),\beta)
\end{equation}
Recalling that $N_{t}\simeq 3\times 9^t/8$ and $B_{t}=9^t$, we have
\begin{equation}
  \omega(p,\beta)=\frac{8}{27}\ln c(p,\beta) +\frac{1}{9}
\omega(\varphi(p,\beta),\beta)
\end{equation}
This yields the following expression for $\omega$
\begin{equation}
\omega(p,\beta)=\sum_{k=0}^{\infty} \frac{1}{9^k}
\frac{8}{27}\ln c(\varphi^{(k)}(p,\beta))
\end{equation}
As above, $\varphi^{(k)}$ is the $k$th iterate of the function $\varphi$.
We note that the recursion relation for $p$ depends parametrically on
$\beta$, and that the temperature is not renormalized in the process.
This implies that there exists a line of critical points, and that the
critical exponents continuously depend on $\beta$ along this line.
From Eqs.~(\ref{eq:eqOmega2}) and~(\ref{eq:lambda}), we obtain an
expression for $x(p,\beta)$
\begin{equation}
x(p,\beta)=\frac{8p}{3}+p(1-p)\frac{\partial \omega}{\partial p}
\end{equation}
We would like to study our model at fixed $x$; the previous formula
tells us how to choose $p$ to do so, providing the bridge between
canonical and grand canonical ensembles. Fig.~\ref{Fig:xp} shows
various $x(p)$ curves, for different values of $\beta$. From
Fig.~\ref{Fig:phiT}, we can see that $p^*(\beta)$ tends to $1$ for
large $\beta$, as we could have anticipated.  However, the critical
connectivity $x^*(\beta)$ does not tend to $8/3$, see
Fig.~\ref{Fig:xp}.
A crucial observable is the energy per site, that is the number of
redundant bonds per site $n_{red}$; it is computed using the formula:
\begin{equation}
n_{red}=\frac{\partial \omega}{\partial \beta} 
\end{equation}
$n_{red}(p,\beta)$ curves are shown on Fig.~\ref{Fig:nred}, together
with results of Monte Carlo simulations.

\subsection{An intermediate phase?}
From the exact solution of the model provided in the previous
paragraph, it is clear that the introduction of an energy and a finite
temperature does not induce any qualitative change in the phase
diagram: as a function of the connectivity, there are only two phases,
floppy and rigid, separated by a phase transition at a critical
connectivity $x^*(\beta)$. Thus, our first conclusion is that in this
model, the "intermediate phase" does not have a real thermodynamical
meaning. It is an important difference with the case of the random
networks~\cite{BBLS_prl,RivoireBarre}. This result seems related to
the strength of the percolation transition, which is first order in
random networks, and very weak on the hierarchical lattices introduced
in this paper.

\begin{figure}
\psfrag{x}{\Large $x$}
\includegraphics[width=14cm,height=14cm]{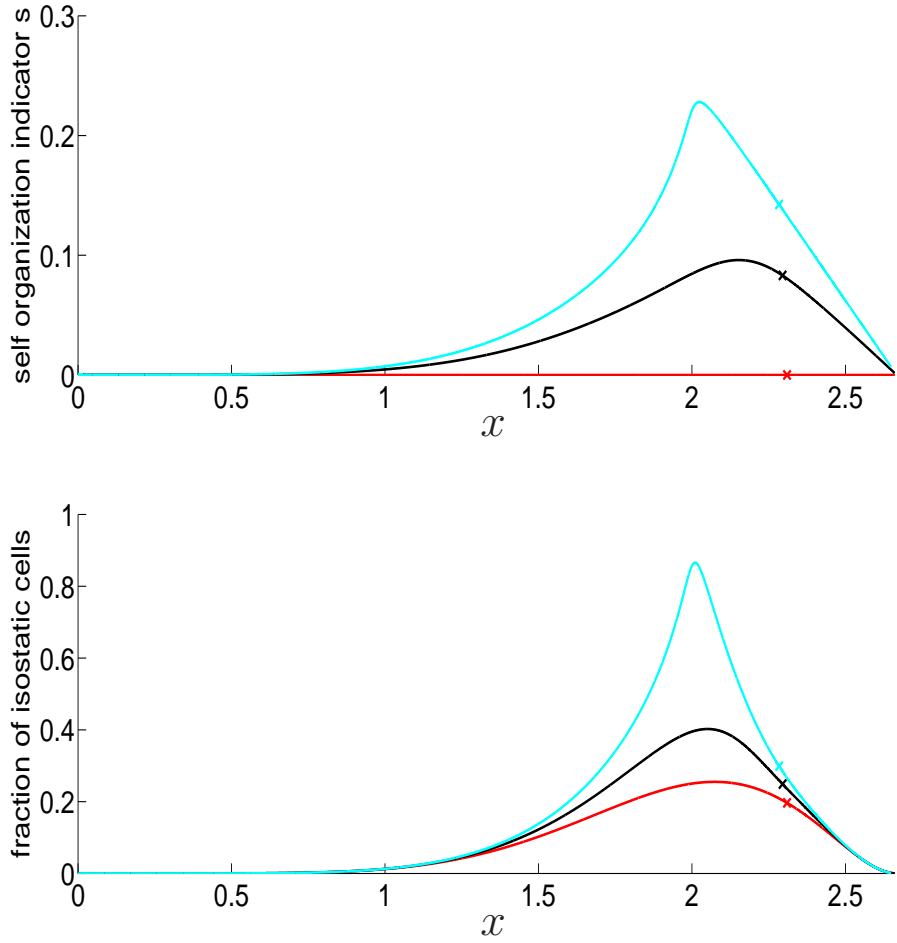}
\caption{(Color online) Top panel: the self organization indicator
  $s=(p-3x/8)/p$, as a function of the mean connectivity $x$. Bottom
  panel: fraction of isostatic elementary cells as a function of the
  mean connectivity.  In both panels, $\beta=0,~1,~5$ from bottom to
  top (respectively red, black and light blue curves), and the crosses
  indicate the critical points. Note that $s$ is identically $0$ when
  $\beta=0$.\label{Fig:selforg}}
\end{figure}

However, looking at the curves on Fig.~\ref{Fig:nred}, it is possible
at low temperatures to distinguish three regions for low enough
temperatures. At low connectivity $x<2$, there is essentially no
redundant bond, and thus no stress in the network; for
$2<x<x^*(\beta)$, there is some stress in the network, but it does
not percolate; finally, for $x>x^*(\beta)$, the stress percolates in
the network. The boundary between the two first regions is not a
thermodynamical transition; rather, it is a cross-over which becomes
sharper as the temperature decreases.  In this cross-over region,
isostatic local structures are favored, and the effect of network
self-organization is the strongest. To illustrate these points, we
have plotted on Fig.~\ref{Fig:selforg} the fraction of isostatic
elementary cells (with exactly $7$ bonds). To define an indicator for
self-organization, we note that without network adaptivity (or
equivalently at infinite temperature), the mean connectivity $x$ is
directly proportional to the parameter $p$, and given by $x=8p/3$; we
may then define a self-organization indicator as $s=(p-3x/8)/p$. $s$
is plotted on Fig.~\ref{Fig:selforg}, and peaks in the cross over
region, just as the fraction of elementary cells. The lower the
temperature, the sharper the peaks get.

Finally, we stress that we find a percolation transition at a critical
density $x_{c}\simeq 2.3$, larger than the Maxwell threshold
$x_{Max}=2$. This contrasts with rigidity models on random graphs, for
which the critical density is usually lower than $x_{Max}$.

\section{Conclusion}
\label{sec:conclusion}
Our first result is the construction of solvable models for rigidity
percolation, beyond trees and random networks. These models display a
critical point at the rigidity percolation threshold, and the critical
exponent for the number of redundant bonds (which corresponds to the
free energy) can be computed exactly. It would be interesting to
compute also the exponent associated with the biggest rigid cluster in
the network. In any case, it is known that these exponents, computed
on such hierarchical networks, cannot be taken as reliable estimates
for the exponents associated to regular lattices. However, these
exponents may converge to the correct ones when one considers
larger and larger cells for the hierarchical
networks~\cite{Reynolds80}. One may then think of a large cell Monte
Carlo renormalization procedure to obtain an estimation of the
critical exponents for 2D generic rigidity percolation. Some words of
caution are in order: it is not clear how to extend to large sizes the
hierarchical networks considered in this paper, nor if they may be a
good approximation of a regular lattice, even for large cell sizes. It
may be necessary to introduce other types of hierarchical lattices.

Our second result concerns the effect of self organization on rigidity
percolation, for these hierarchical models. We define and solve
exactly a model where the network may adapt its structure to avoid
stress, at an entropic cost. At variance with what happens for random
networks, the possibility of self-organization does not introduce a
qualitative change in the phase diagram. In particular, there is no
true thermodynamic ``intermediate phase''. It may still be possible to
define an ``intermediate region'', where the network is close to be
isostatic, and the stress does not percolate, although it is present
locally. It would be interesting to compare the properties of this
``intermediate region'' with the picture of the intermediate phase
drawn from a Size Increasing Cluster
Approximation~\cite{MicoulautEPL02}. From the modeling point of view,
we note that a scenario where sharp phase transitions are replaced by
smoother cross-overs may still be compatible with the experiments on
network glasses. 
 
All our analysis relies on the combinatorial description of 2D generic
rigidity. It should be possible to study similar hierarchical models of 3D
rigidity with bond bending constraints, using the associated
combinatorial description~\cite{pebble3d}. It is not clear however how
to generalize the technique to 3D networks with central forces only,
as there is no simple combinatorial description in that
case~(see~\cite{Chubynsky07} for a recent discussion).

\paragraph*{Acknowledgments ---} The author warmly thanks M.
Chubynsky, for making available to him his implementation of the
Pebble Game algorithm, and acknowledges useful discussions with S.
Redner.


\begin{thebibliography}{99}

\bibitem{Phillips79} J.C.~Phillips, \emph{J.~Non-Cryst.~Solids} {\bf 34}, 153
  (1979).

\bibitem{Thorpe83} M.F.~Thorpe, \emph{J.~Non-Cryst.~Solids} {\bf 57}, 355
  (1983). 

\bibitem{LatvaPRE01} M. Latva-Kokko, J.Timonen {\emph Phys. Rev. E} 
{\bf 64}, 066117 (2001).

\bibitem{HeadPRL03} D.A. Head, A. J. Levine, and F.C.MacKintosh 
{\emph Phys. Rev. Lett.} {\bf 91}, 108102 (2003).
  
\bibitem{Hendrickson}  D.J.~Jacobs and B. Hendrickson \emph{J. Comp. Phys.}
{\bf 137}, 346 (1997).

\bibitem{PebbleJacobs} D.J.~Jacobs and M.F.~Thorpe,
  \emph{Phys. Rev. Lett.} {\bf 75}, 4051 (1995).

\bibitem{PebbleMoukarzel} C. Moukarzel and P.M.~Duxbury,
  \emph{Phys. Rev. Lett.} {\bf 75}, 4055 (1995).

\bibitem{Laman} G. Laman, \emph{J. Eng. Math.} {\bf 4}, 331 (1970).

\bibitem{PebblePRE}D.J. Jacobs and M.F. Thorpe, \emph{Phys. Rev E} {\bf 53},
3682 (1996).

\bibitem{MoukarzelRBM99} P.M.~Duxbury, D.J.~Jacobs, M.F.~Thorpe and
  C.~Moukarzel, \emph{Phys. Rev. E} {\bf 59}, 2084 (1999).

\bibitem{Obukhov} S.P. Obukhov, \emph{Phys. Rev. Lett.} {\bf 74},
4472 (1995).

\bibitem{DuxburyPRE97} C. Moukarzel, P.M. Duxbury and P.L. Leath,
\emph{Phys. Rev. E} {\bf 55}, 5800 (1997).

\bibitem{ThorpeBook99} M.F.~Thorpe, D.J.~Jacobs, M.V.~Chubinsky
and  A.J.~Rader, in Rigidity Theory and Applications, Ed. by
M.F. Thorpe and P.M. Duxbury (Kluwer Academic/Plenum Publishers, New
York, 1999).

\bibitem{ChubinskyThesis} M.V.~Chubinsky, PhD Thesis, Michigan State
  University (2003).

\bibitem{BBLSjsp} J. Barr\'e, A. Bishop, T. Lookman and A. Saxena, 
\emph{J. Stat. Phys.} {\bf 118}, 1057 (2005).

\bibitem{Sahimi85} S.~Feng and M.~Sahimi, \emph{Phys. Rev. B} {\bf 31},
  1671 (1985). 

\bibitem{Sahimi92} M.A. Knackstedt and M. Sahimi, \emph{J.~Stat.~Phys.}
{\bf 69}, 887 (1992).

\bibitem{ThorpeJNCS00} M.F.~Thorpe, D.J.~Jacobs, M.V.~Chubinsky and
  J.C.~Phillips, \emph{J. Non-Cryst. Sol.} {\bf 266-269}, 859 (2000). 

\bibitem{BoolchandPRB00} D.~Selvanathan, W.J. Bresser and P.
  Boolchand, \emph{Phys. Rev.~B} {\bf 61}, 15061 (2000). 

\bibitem{BoolchandEPL00} Y. Wang, P.~Boolchand and M. Micoulaut,
  \emph{Europhys. Lett.}  {\bf 52}, 633 (2000).

\bibitem{BoolchandJNCS01} P.~Boolchand, X.~Feng and W.J.~Bresser
  \emph{J. Non Cryst. Solids}  {\bf 293-295}, 348 (2001).

\bibitem{MicoulautEPL02} M. Micoulaut, \emph{Europhys. Lett.} {\bf
58}, 830 (2002).

\bibitem{MicoulautPhillips} M.~Micoulaut and J.C.~Phillips,
\emph{Phys. Rev. B} {\bf 67}, 104204 (2003).

\bibitem{BBLS_prl} J. Barr\'e, A.R. Bishop, T. Lookman and A. Saxena, 
\emph{Phys. Rev. Lett.} {\bf 94}, 208701 (2005).

\bibitem{RivoireBarre} O. Rivoire and J. Barr\'e \emph{Phys. Rev. Lett.}
{\bf 97}, 148701 (2006).

\bibitem{ChubynskyBriere06} M.V. Chubynsky, M.-A. Bri\`ere, N.
Mousseau, \emph{Phys. Rev E} {\bf 74}, 016116 (2006).

\bibitem{ChubynskyBriere06b} M.-A. Bri\`ere, M.V. Chubynsky, N.
Mousseau, \emph{Phys. Rev. E} {\bf 75}, 056108 (2007).

\bibitem{Sartbaeva07} A. Sartbaeva, S.A. Wells, A. Huerta and M.F. Thorpe,
\emph{Phys. Rev. B} {\bf 75}, 224204 (2007).

\bibitem{MicoulautJNCS07} M.~Micoulaut and J.C.~Phillips, 
\emph{J. Non Cryst. Solids} {\bf 353}, 1732 (2007).

\bibitem{Chubynsky08} M. Chubynsky, preprint; cond-mat arXiv:0807.2887.

\bibitem{Ostlund} A.N. Berker and S. Ostlund \emph{J. Phys C} {\bf
    12}, 4961 (1979).

\bibitem{Kaufman81} M. Kaufman and R.B. Griffiths
  \emph{Phys. Rev. B} {\bf 24}, 496 (1981).

\bibitem{Kaufman82}  R.B. Griffiths and M. Kaufman
  \emph{Phys. Rev. B} {\bf 26}, 5022 (1982).

\bibitem{Reynolds80} P.J. Reynolds, H.E. Stanley and W. Klein
  \emph{Phys. Rev. B} {\bf 21}, 1223 (1980).

\bibitem{pebble3d} D.J.~Jacobs, \emph{J. Phys. A: Math. Gen.} {\bf
  31}, 6653 (1998).

\bibitem{Chubynsky07} M.V. Chubynsky and M.F. Thorpe \emph{Phys. Rev.
    E} {\bf 76}, 041135 (2007).

\end{thebibliography}
\end{document}